\documentstyle{article}

\newcommand{\EQ}{\begin{equation}}
\newcommand{\EN}{\end{equation}}
\newcommand{\bea}{\begin{eqnarray}}
\newcommand{\ena}{\end{eqnarray}}
\newcommand{\vs}[1]{\vspace{#1 mm}}

\renewcommand{\b}{\beta}
\renewcommand{\c}{\gamma}

\newcommand{\e}{\epsilon}
\renewcommand{\t}{\theta}
\newcommand{\tb}{{\bar \theta}}
\newcommand{\shalf}{\frac{1}{2}}
\newcommand{\pa}{\partial}
\newcommand{\nn}{\nonumber \\}

\begin{document}

\topmargin 0pt
\oddsidemargin 5mm

\renewcommand{\Im}{{\rm Im}\,}
\newcommand{\NP}[1]{Nucl.\ Phys.\ {\bf #1}}
\newcommand{\PL}[1]{Phys.\ Lett.\ {\bf #1}}
\newcommand{\CMP}[1]{Comm.\ Math.\ Phys.\ {\bf #1}}
\newcommand{\PR}[1]{Phys.\ Rev.\ {\bf #1}}
\newcommand{\PRL}[1]{Phys.\ Rev.\ Lett.\ {\bf #1}}
\newcommand{\PTP}[1]{Prog.\ Theor.\ Phys.\ {\bf #1}}
\newcommand{\PTPS}[1]{Prog.\ Theor.\ Phys.\ Suppl.\ {\bf #1}}
\newcommand{\MPL}[1]{Mod.\ Phys.\ Lett.\ {\bf #1}}
\newcommand{\IJMP}[1]{Int.\ Jour.\ Mod.\ Phys.\ {\bf #1}}

\begin{titlepage}
\setcounter{page}{0}
\begin{flushright}
KCL-TH-94-6,
OU-HET 189\\
May 1994\\
hep-th/9405144
\end{flushright}

\vs{10}
\begin{center}
{\large\bf Embeddings for Non-Critical Superstrings}

\vs{15}
{\large Nathan Berkovits\footnote{e-mail: udah101@kcl.ac.uk}}\\
{\em Math Dept., King's College,
Strand, London, WC2R 2LS, United Kingdom}

\vs{8}
{\large Nobuyoshi Ohta\footnote{e-mail: ohta@fuji.wani.osaka-u.ac.jp}}\\
{\em Department of Physics,
Osaka University, Toyonaka, Osaka 560, Japan}\\

\end{center}

\vs{15}
\centerline{{\bf{Abstract}}}

It was previously shown that at critical central charge, $N$-extended
superstrings can be embedded in $(N+1)$-extended superstrings. In other
words, $(N=0,c=26)\to (N=1,c=15)\to (N=2,c=6)\to (N=3,c=0) \to (N=4,c=0) $.
In this paper, we show that similar embeddings are also possible for
$N$-extended superstrings at non-critical central charge.
For any $x$, the embedding is $(N=0,c=26+x) \to (N=1,c=15+x) \to (N=2,c=6+x)
\to (N=3,c=x) \to (N=4,c=x)$. As was conjectured by Vafa, the $(N=2,c=9) \to
(N=3,c=3)$ embedding can be used to prove that $N=0$ topological strings
are special vaccua of N=1 topological strings.

\end{titlepage}
\newpage
\renewcommand{\thefootnote}{\arabic{footnote}}
\setcounter{footnote}{0}

It has recently been noticed \cite{R,B,BV,OP} that starting with a critical
$N$-extended stress tensor and with a twisted set of $N$-extended
``ghosts'', it is possible to construct a critical $(N+1)$-extended
stress tensor (this procedure differs from that of refs.~\cite{GS,BLN}
in not requiring the existence of a $U(1)$ current).
The BRST charge
constructed out of this $(N+1)$-extended stress tensor has the same
cohomology as the BRST charge constructed out of the original
$N$-extended stress tensor~\cite{FO,OP}. Furthermore,
it was shown for the $(N=0,c=26) \to (N=1,c=15)$ and
$(N=1,c=15) \to (N=2,c=6)$
embeddings that if the $(N+1)$-extended stress tensor corresponds to
the matter sector of a critical $(N+1)$-extended string, the
$(N+1)$-extended prescription for calculating scattering amplitudes
gives the same result as the original $N$-extended prescription~\cite{BV}.

In this paper, it will be shown that for $N$-extended stress tensors
with non-critical central charge, it is also possible to use a twisted
set of $N$-extended ghosts to construct an $(N+1)$-extended stress tensor.
Although the $(N+1)$-extended stress tensor has non-critical central
charge, the difference between the central charges of the $N$-extended
and $(N+1)$-extended stress tensors is always equal to the difference
of the critical central charges. In other words, we will describe
the following embeddings:
\EQ
(N=0, c=26+x) \to
(N=1, c=15+x) \to
(N=2, c=6+x) \to
(N=3, c=x)\to (N=4, c=x),
\EN
where $x$ is arbitrary.
Note that as in the critical case, these embeddings do not require the
existence of a $U(1)$ current.

There are various interesting features of these non-critical embeddings
that will be described. For $N=0 \to N=1$ and $N=1 \to N=2$, it is
possible to generalize the critical embeddings of
ref.~\cite{BV} by simply
changing some of the coefficients in the stress tensor. However, for
$N=2 \to N=3$ and $N=3 \to N=4$, the non-critical $(N+1)$-extended
stress tensor requires an infinite number of terms, unlike the critical
case described in ref.~\cite{OP}.
Surprisingly, it is not difficult to determine the
explicit structure of these infinite terms.

Another interesting point is that there are two very
different ways to construct the
$N=1 \to N=2$ non-critical
embedding. The first construction has only a finite
number of terms but requires bosonization of the twisted ghosts.
The second construction shares the structure of the the other $N\to N+1$
embeddings but requires an infinite number of terms for the critical,
as well as the non-critical case.

As an important application of our non-critical embeddings, it was
conjectured by Vafa that our result on the $(N=2,c=9) \to (N=3,c=3)$
embedding can be used to prove the equivalence of the scattering
amplitudes of $N=0$ and $N=1$ topological strings.
This will be discussed briefly after presenting our non-critical
embeddings, and will be described in more detail in ref.~\cite{V}.

We begin this paper by giving a general procedure for turning
a critical embedding into a non-critical embedding, and will then give
explicit expressions for the non-critical embeddings.
Except for the
$N=0 \to N=1$ embedding (which was also found independently by Amit
Giveon), these embeddings were constructed with the aid of the computer
program of \cite{TK}. The $N=1\to N=2$ non-critical
embedding will be descibed after the others because it has some
fundamentally different features.
We will conclude with some comments on the significance of our results.

\vskip 20pt
The $N\to (N+1)$ critical embedding for $N\geq 2$ is~\cite{OP,UP}
\EQ
T_{crit}=T_m^{crit} + T_g ,\quad G_{crit} =B+ j ,
\label{creal}
\EN
where all objects are
$N$-extended superfields which depend on $z$ and
on $N$ grassmann parameters, $T_{crit}$ and $G_{crit}$ are the
components of the critical
$(N+1)$-extended stress tensor, $T_m^{crit}$ is a critical $N$-extended
stress-tensor, $B$ and $C$ are the twisted $N$-extended ghosts,
$T_g$ is the $N$-extended stress tensor for the twisted ghosts,
and $j$ is the integrand of the $N$-extended BRST charge.
For the $N=0\to N=1$ critical embedding, there exist total
derivative correction terms to
$T_{crit}$ and $G_{crit}$, however
these correction terms do not affect
the construction described below. The $N=1\to N=2$ critical
embedding described in refs.~\cite{B,BV} is fundamentally different
from the other cases and will be discussed seperately.

To construct a non-critical $N\to(N+1)$ embedding from the critical
embedding of eq.~(\ref{creal}),
one first introduces a new $(N+1)$-extended stress tensor,
$\hat T$ and $\hat G$, where $\hat T$ and $\hat G$ describe
a $c=x$ representation which is unrelated to $T_m^{crit}$ and $T_g$.
It is clear that if $T_{crit}$ and $G_{crit}$ describe an $(N+1)$-extended
stress tensor with $c=c_{crit}$, then $T'=T_{crit}+\hat T$ and
$G'=G_{crit}+\hat G$
describe an $(N+1)$-extended stress tensor with $c=c_{crit} +x$.

The next step is to find a similarity transformation that removes
all explicit dependence of $\hat G$ from the non-critical stress tensor
described by $T'$ and $G'$.
This can be done inductively by first defining
\EQ
T''= e^{-\int  C \hat G}~ T'~ e^{\int  C\hat G},\quad
G''= e^{-\int  C \hat G} ~G'~ e^{\int  C\hat G}
\label{sim1}
\EN
where $\int$ signifies a super-integration over
$z$ and over the $N$ grassmann
parameters.

It is easy to use eq.~(2) to
check that $T''=T'$ and $ G''=
G' -\hat G +C\hat T + ...$, where $...$ signifies terms with at
least ghost-number 2 (the ghost number operator is $\int CB$).
Therefore, after the similarity transformation,
\EQ
T''=(T_m^{crit} + \hat T) +T_g,\quad G''= B + (j +C\hat T) +...
\label{sim2}
\EN
Let $Y$ be all terms in $...$ with ghost-number 2 containing either
$\hat T$ or $\hat G$.
Since the OPE of $G''$ with $G''$ has only ghost-number zero terms
($T''$ has ghost-number zero), the symmetrized
OPE of $B$ with $Y$ must be non-singular.
Therefore,
\EQ
[ \int CY~, ~B]~ =  ~-[B~,~ \int C] ~Y~- [\int CB~,~ Y] ~
= ~-3 Y.
\EN

So after performing the similarity transformation,
\EQ
T'''= e^{{1\over 3}\int C Y}~ T''~ e^{-{1\over 3}\int CY}=T'', \quad
G'''= e^{{1\over 3}\int C Y}~ G''~ e^{-{1\over 3}\int CY}=G'' -Y+...,
\label{sim3}
\EN
all terms containing $\hat G$ have at least ghost-number 3.
By repeating this inductive procedure, one can construct the
$(N+1)$-extended stress tensor with central charge $c=c_{crit}+x$:
\EQ
T=(T_m^{crit} +\hat T) + T_g,\quad G=B +(j+C\hat T) +...,
\EN
where the $...$ depends only on the $N$-extended ghosts. Since
$T_m^{crit}$ and $\hat T$ only appear in the combination
$T_m^{crit} +\hat T$, this
combination
can be replaced by an arbitrary non-critical $N$-extended stress tensor
$T_m$ which is then embedded into the non-critical $(N+1)$-extended stress
tensor $T$.

The explicit non-critical $N \to (N+1)$ embeddings constructed using
this inductive procedure are as follows:
\bea
N=0 \to N=1:\quad
T &=&  T_m - \frac{3}{2}B\pa C- \shalf \pa BC
 + \shalf \pa^2(C \pa C), \nn
 G &=& B + C T_m + BC \pa C
 - \frac{x}{24}C \pa C \pa^2 C+ \frac{15+x}{6}\pa^2 C,
\label{real1}
\ena
which satisfy the $N=1$ OPE:
\bea
T(z) T(w) &\sim&
\frac{\frac{1}{2}(15+x)}{(z-w)^4} + \frac{2 T}{(z-w)^2}
 + \frac{\pa T}{z-w}, \nn
T(z) G(w) &\sim&
\frac{\frac{3}{2} G}{(z-w)^2} + \frac{\pa G}{z-w}, \nn
G(z) G(w) &\sim&
\frac{\frac{2}{3}(15+x)}{(z-w)^3} + \frac{2 T}{z-w}.
\ena
\bea
N=2 \to N=3:\quad
T &=&  T_m -
\shalf\pa(BC)+(DC)({\bar D}B)+({\bar D}C)(DB),\nn
G &=& B + C T_m + C(DC)({\bar D}B)+C({\bar D}C)(DB)
 - B({\bar D}C)(DC) - \left(\frac{x}{3}\right)[D,{\bar D}]C \nn
&+& \sum_{n=1}^\infty \frac{x}{6} \left[ -C (DC)({\bar D}\pa C)
 +C({\bar D}C)(D\pa C) - 2(DC)({\bar D}C)[D,{\bar D}]C \right.\nn
&-& \left.(2n-1)C(D{\bar D}C)({\bar D}DC)\right]\left[ (DC)({\bar D}
C)\right]^{n-1},
\label{real3}
\ena
where $D= \pa_{\t}-\shalf\tb\pa_z$ and
${\bar D} = \pa_{\tb}-\shalf\t\pa_z$ are the usual $N=2$ fermionic
derivatives, all fields are $N=2$ superfields,
$C(Z_1)B(Z_2) \sim \t_{12}\tb_{12}/z_{12}$, $\t_{12}\equiv
\t_1 -\t_2$, $z_{12} \equiv
z_1 - z_2 + \shalf (\t_1\tb_2 + \tb_1\t_2)$, and the $N=3$ OPE is
\bea
T(Z_1) T(Z_2) &\sim&
\frac{\frac{1}{3}x +\t_{12}\tb_{12} T}{z_{12}^2}
 + \frac{-\t_{12} DT +
\tb_{12}{\bar D} T + \t_{12}\tb_{12}\pa T}{z_{12}},\nn
T(Z_1) G(Z_2) &\sim& \shalf
\frac{\t_{12}\tb_{12}}{z_{12}^2} G
 + \frac{-\t_{12} DG +\tb_{12}{\bar D}G +
\t_{12}\tb_{12}\pa G}{z_{12}},\nn
 G(Z_1) G(Z_2) &\sim& \frac{\frac{4}{3}x+
\t_{12}\tb_{12}2T}{z_{12}^2}.
\ena

\vskip 15pt
$N=3 \to N=4$:
\bea
 T &=& T_m -\shalf C\pa B +\shalf (D_l C)(D_l B), \nn
G &=& B + C T_m  -
 \frac{1}{4}(D_l C)^2 B +\shalf C (D_l C)(D_l B) +
 \shalf C \pa C B - \frac{x}{36}\int^z \e^{ijk}D_iD_jD_k C \nn
&-& \sum_{n=1}^\infty \frac{x \e^{ijk}}{36(2n+1)4^n}
 \left[ \int^z D_iD_jD_k \left[ C((D_l C)^2)^{n} \right]
 - 3 C(D_i D_j C)(D_k C)((D_l C)^2)^{(n-1)} \right],
\label{real4}
\ena
where the sum over $i,j,k,l=1,2,3$ is understood, $\int^z$ stands for
an ordinary integration up to the point $z$,
$D_i=\pa_{\t^i} +\t^i\pa_z$ are the usual N=3 fermionic derivatives,
$C(Z_1)B(Z_2) \sim \t^1_{12}\t^2_{12}\t^3_{12}/z_{12}$, $z_{12}
\equiv z_1 - z_2 - \t^i_1\t^i_2$, and the $N=4$ OPE is
\bea
T(Z_1) T(Z_2) &\sim& \frac{ -\frac{1}{12}x
 +\frac{1}{4}\e_{ijk}\t^i_{12}\t^j_{12} D_k T
 + \t^1_{12}\t^2_{12}\t^3_{12}\pa T}{z_{12}}
 + \shalf \frac{\t^1_{12}\t^2_{12}\t^3_{12} T}{z^2_{12}}, \nn
T(Z_1) G(Z_2) &\sim& \frac{\frac{1}{4}\e_{ijk}\t^i_{12}\t^j_{12} D_k G
 + \t^1_{12}\t^2_{12}\t^3_{12}\pa G}{z_{12}}, \nn
G(Z_1) G(Z_2) &\sim& \frac{x}{3} \ln z_{12}
 + \frac{ \t^1_{12}\t^2_{12}\t^3_{12}}{z_{12}} 2 T.
\ena
Note that the $\int^z$ in $G$ causes the lowest component of the
non-critical N=4 stress-tensor to be non-local,
however this is expected since the OPE of the lowest component with
itself goes like $\log (z_1-z_2)$.

\vskip 20pt

The critical $N=1\to N=2$ embedding described in ref.~\cite{BV}
has the form
\EQ
T_{crit}= T_m^{crit} +T_g, \quad G^-_{crit}=b, \quad  G^+_{crit}
= j, \quad J_{crit}=cb-\xi\eta,
\label{creal2}
\EN
where $T_m^{crit}$ is a $c=15$ stress-tensor, $b$ and $c$ are the fermionic
Virasoro ghosts, $\b=\pa \xi e^{-\phi}$ and $\c=\eta e^{\phi}$
are the bosonic super-Virasoro ghosts, $T_g$ is the $c=-9$ stress-tensor
for the twisted ghosts, and $j$ is the $N=1$ BRST current (it
includes total derivative correction terms). Note
that this embedding requires manifest $N=1$ supersymmetry to be broken
and also requires bosonization of the super-Virasoro ghosts.

A non-critical $N=1\to N=2$ embedding can be obtained using a similar
procedure as before. One first introduces a new $N=2$ stress tensor,
$(\hat T, \hat G^-, \hat G^+, \hat J)$, which has central charge $c=x$
and which is unrelated to $T_m^{crit}$ and $T_g$. Adding this $N=2$
stress-tensor to the critical stress tensor of eq.~(\ref{creal2})
produces an $N=2$ stress-tensor,
$(T=T_{crit}+\hat T,
G^-=G^-_{crit}+\hat G^-,
G^+=G^+_{crit}+\hat G^+,
J=J_{crit}+\hat J)$,
with
$c=6+x$. One then has to find a similarity transformation that removes
all explicit dependence on $\hat G^+ - \hat G^-$ and $\hat J$.

The resulting non-critical $N=1\to N=2$ embedding is:
\bea
T &=&  T_m - \frac{3}{2}b\pa c- \shalf \pa c b
 - \frac{3}{2}\eta\pa\xi - \shalf \pa\eta\xi
 - \shalf (\pa\phi)^2 - \pa^2\phi, \nn
 G^+ &=& \eta e^\phi G_m + c T_m - \shalf c(\pa\phi)^2
 - \frac{12+x}{12}c \pa^2\phi - b\eta\pa\eta e^{2\phi}
 + \frac{6+x}{6} \pa c\xi \eta \nn
&+& \frac{24+x}{12} c\pa\xi \eta
 + \frac{12+x}{12} c \xi \pa\eta + c\pa cb
 + \frac{6+x}{6} \pa^2 c- \frac{x}{6}\pa c\pa\phi,\nn
G^- &=& b, \nn
J &=& cb - \frac{6+x}{6}\xi\eta +\frac{x}{6}\pa\phi,
\label{real2}
\ena
where the $N=2$ OPE is
\bea
T(z) T(w) &\sim&
\frac{\frac{1}{2}(6+x)}{(z-w)^4} + \frac{2 T(w)}{(z-w)^2}
 + \frac{\pa  T(w)}{z-w}, \nn
 T(z) G^\pm(w)
&\sim& \frac{\frac{3}{2} G^\pm(w)}{(z-w)^2}
 + \frac{\pa G^\pm(w)}{z-w}, \nn
 T(z) J(w) &\sim&
\frac{ J(w)}{(z-w)^2} + \frac{\pa J(w)}{z-w}, \nn
G^+(z) G^-(w) &\sim&
\frac{\frac{1}{3}(6+x)}{(z-w)^3} + \frac{ J(w)}{(z-w)^2}
 + \frac{ T(w) + \shalf \pa  J(w)}{z-w},\nn
 J(z)G^\pm(w)
&\sim& \frac{\pm  G^\pm(w)}{z-w}, \nn
 J(z) J(w) &\sim& \frac{\frac{1}{3}(6+x)}{(z-w)^2}.
\ena

It is natural to ask if there also exists an $N=1\to N=2$ embedding
which preserves manifest $N=1$ supersymmetry and does not require
bosonization. In fact, we have found such an embedding
using the computer and inspired guesswork, and have explicitly
checked that the $N=2$ OPE's are satisfied up to terms with
ghost-number greater than 11.
This second type of $N=1\to N=2$ embedding can be
written in $N=1$ superfield notation as:
\bea
 T &=& T_m + \shalf (DB)(DC) -B \pa C - \shalf \pa B C
+ \sum_{n=1}^{\infty} \frac{1}{4^n}\pa\left[ C(D\pa C)(DC)^{2n-2}
 \right] , \nn
G &=& B + C T_m + \shalf B\pa C C -\frac{1}{4}(DC)^2 B
 +\shalf DC C DB - \frac{6+x}{6} D\pa C  \nn
&+& \sum_{n=1}^{\infty} \frac{1}{4^n} [ (1+(n-1)\frac{6+x}{3})
 C\pa C(D\pa C) (DC)^{2n-2} \nn
&& + \frac{6+x}{6} C \pa^2 C(DC)^{2n-1}
 - \frac{6+x}{6} D\pa C(DC)^{2n} ],
\label{real21}
\ena
where $D=\pa_{\t}+\t\pa_z$ is the $N=1$ fermionic derivative and
the $N=2$ OPE is
\bea
T(Z_1)T(Z_2) &\sim& \frac{\frac{1}{6}(6+x)}{z_{12}^3}
 + \frac{3}{2}\frac{\t_{12}T}{z_{12}^2}
 + \frac{\shalf DT + \t_{12}\pa T}{z_{12}},\nn
T(Z_1)G(Z_2) &\sim&  \frac{\t_{12}G}{z_{12}^2}
 + \frac{\shalf DG + \t_{12}\pa G}{z_{12}},\nn
G(Z_1)G(Z_2) &\sim& \frac{\frac{1}{3}(6+x)}{z_{12}^2}
 + \frac{\t_{12}2T}{z_{12}}.
\ena
Note that the embedding requires
an infinite number of terms even when $c=c_{crit}=6$ (i.e. $x=0$).

Although the expressions for $G$ in eqs.~(\ref{real3}),
(\ref{real4}), and (\ref{real21}),
contain terms of arbitrarily high ghost number,
the structure of these terms is fixed by the requirement that
they have the correct OPE with $T$. It is easy to show
that at each ghost-number, this completely determines the terms up
to an overall coefficient. The overall coefficient at each ghost number
can be determined by analyzing the OPE of $G$ with
$G$.

\vskip 20pt

Because these embeddings are for non-critical string theories, it is not
straightforward to compare scattering amplitudes using the $N$-extended
and $(N+1)$-extended prescriptions. However there is one special case
where the non-critical embedding does allow a comparison of scattering
amplitudes. If one starts with a $(N=2,c=9)$ matter sector and twists
the $N=2$ stress tensor in the usual way, $N=0$ topological string
amplitudes can be calculated by pretending it is a bosonic string theory
with the $b$ ghosts replaced by the spin 2 fermionic generator.
Similarly, if one starts with a $(N=3,c=3)$ matter sector and twists
with respect to one of the $SO(3)$ generators, $N=1$ topological
amplitudes can be calculated by pretending it is an $N=1$ superstring
with the $b$ and $\beta$ ghosts replaced by spin 2 and spin 3/2
generators~\cite{BLN}. So given a $(N=2,c=9)$ matter sector, one can
either calculate $N=0$ topological amplitudes, or use the non-critical
$N=2 \to N=3$ embedding to construct a $(N=3,c=3)$ matter sector and
calculate $N=1$ topological amplitudes.

It was conjectured by Cumrun
Vafa that these two scattering amplitudes coincide, and it is
straightforward to use the properties
of the non-critical embedding
to prove his conjecture correct. The proof of equivalence for
$N=0$ and $N=1$ topological amplitudes is very similar to
the proof of ref.~\cite{BV} for $N=0$ and $N=1$ ordinary amplitudes.
As will be discussed in more detail in ref.~\cite{V}, one needs to
insert picture-changing operators in the $N=1$ prescription
(which for topological strings are the products of
spin 3/2 fermionic generators and the delta-functions of
spin 3/2 bosonic generators), as well as beltrami differentials sewn with
the spin 2 fermionic generators.
If the $(N=3,c=3)$ matter sector comes from
the non-critical $N=2\to N=3$ embedding described in eq.~(\ref{real3}),
these spin 3/2 generators depend linearly on the twisted ghosts which
were added to the original $(N=2,c=9)$ matter sector.
It is easy to check that these picture-changing operator insertions
absorb the zero modes of the twisted ghosts and
that the non-zero modes of the twisted ghosts
cancel each other out. So after integrating over
the twisted ghosts, only the spin 2 fermionic generators
remain in the functional integral, and
the $N=1$ topological prescription reduces to the
original $N=0$ topological prescription.

\vskip 20pt

In this paper, we have shown that the embeddings of critical superstrings
found in earlier works can be generalized to embeddings of non-critical
superstrings. Because our constructions of these new embeddings are
rather complicated, it is natural to ask if there is an underlying
principle that guarantees their existence. Certainly at the classical
level, it is always possible to embed a system with less symmetry into
a system with more symmetry by simply adding ``artificial'' gauge
degrees of freedom. However at the quantum level, things are not so
simple.

For example, it appears that without the presence of a $U(1)$ current,
it is only possible to embed the $N$-extended string into an
$(N+1)$-extended string if the difference of the central charges is
equal to the difference of the critical central charges. Although this
allows the $N=0$ topological string to be embedded into the $N=1$
topological string, a similar embedding is not possible for the $N=1$
topological string into the $N=2$ topological string. The reason is
that an $N=2$ topological
string comes from twisting an $N=4$ superstring with vanishing central
charge (the central charge is three times the anomaly of the ghost-number
current, which is zero for the $N=2$ string). However the $N=3 \to N=4$
embedding of eq.~(\ref{real4}) maps the $(N=3,c=3)$ superstring
into the
$(N=4,c=3)$ superstring, rather than the desired $(N=4,c=0)$ superstring.
It would be interesting to learn if there is another $N=3 \to N=4$
non-critical embedding which embeds the $N=1$ topological string into
the $N=2$ topological string, or if there is a fundamental obstruction
to such an embedding.

Another interesting question is why the $N=1 \to N=2$ embedding looks
so different from all other $N \to N+1$ embeddings. As was shown in
eqs.~(\ref{real2}) and (\ref{real21}), it is only possible to embed
$N=1$ into $N=2$ using a finite
number of terms if one bosonizes the twisted $(\b,\c)$ ghosts.
Without such a bosonization, even the critical embedding requires an
infinite number of terms. The $N=1 \to N=2$ embedding is of special
significance because it allows the Ramond-Neveu-Schwarz description of the
ten-dimensional superstring to be related to the manifestly
spacetime-supersymmetric Green-Schwarz description~\cite{B}.

\vskip 20pt
Acknowledgements: NB would like to thank Jos\'e Figueroa-O'Farrill,
Amit Giveon, Chris Hull, and Cumrun
Vafa for useful discussions, and the SERC for financial support.
NO would like to thank Fiorenzo Bastianelli, Hiroshi Kunitomo and
Jens Lyng Petersen for valuable discussions and Paolo Di Vecchia for
hospitality at NORDITA where part of this work was done.

\end{document}